The tail risks of FX return distributions: a comparison of the returns associated with limit orders and market orders

By

John Cotter and Kevin Dowd[*]

Abstract

This paper measures and compares the tail risks of limit and market orders using Extreme Value Theory. The analysis examines realised tail outcomes using the Dealing 2000-2 electronic broking system based on completed transactions rather than the more common analysis of indicative quotes. In general, limit and market orders exhibit broadly similar tail behaviour, but limit orders have significantly heavier tails and larger tail quantiles than market orders.

May 18, 2007

[*] John Cotter is at the Centre for Financial Markets, School of Business, University College Dublin, Carysfort Avenue, Blackrock, Co. Dublin, Ireland; email: john.cotter@ucd.ie. Kevin Dowd is at the Centre for Risk and Insurance Studies, Nottingham University Business School, Jubilee Campus, Nottingham NG8 1BB, UK; email: Kevin.Dowd@nottingham.ac.uk. The authors would like to thank Charles Goodhart, Jon Danielsson and Richard Payne for helpful comments on an earlier draft, but are responsible for any remaining errors. Cotter's contribution to the study has been supported by a University College Dublin School of Business research grant.

# 1 Introduction:

Trades in financial markets can be broadly divided into two types: limit orders, which are orders to buy or sell at some pre-specified price, and market orders, which are orders to buy or sell immediately at going market prices. Over the years there has been extensive discussion of the relative merits of these two types of order. A market order has the advantage of ensuring a trade, but only at the going market price. On the other hand, a limit order does not necessarily become active, but if it does, it generally gives the trader a better price than a market order. For example, Harris (1998) notes that informed traders prefer using market orders in highly liquid markets, and Danielsson and Payne (2002a) note that in periods of high trading activity market order traders have better information than limit order traders. In contrast, limit orders can be more attractive for order submission in the case of asymmetric information (Handa and Schwartz, 1996), and non-informed traders might prefer limit orders depending on their trading objectives.

A key question is whether limit and market orders have similar risk and return characteristics. This raises the issue of the how the return distributions of these two types of order might compare with each other. This is important, because many microstructure models require that the data exhibit certain distributional characteristics (see Lyons, 2001; and Hasbrouck, 2007; for a discussion on foreign exchange and equity characteristics). For example, Gabaix et al (2003) present results suggesting that the distributions of return series have only the first three moments defined, and this would suggest that the kurtosis is infinite. There is also some dispute over whether return distributions even have a finite second moment: although most scholars accept that they do, this point of view is controversial, and Mittnik et alia (1998, 2000) make a good argument against it. This is an important issue because many models require the variance of price changes to be finite: an example is Roll's (1984) model of bid, ask and transaction prices. At the very least, we cannot therefore be sure how many moments actually exist, and this has major implications for the models we can apply.

Leaving aside these issues, it is also true to say that a commonly observed stylized fact for the empirical distribution of exchange rate returns is their heavy-tails: these have been well-documented for different frequencies and under different institutional frameworks (see, e.g., Cotter, 2005). We seek to investigate these heavy tails far out into the tails using Extreme Value Theory (EVT). The limiting EVT distribution nests a number of distributions and allows for analysis of tail shape where distinctions can be made based on the empirical features of order type. The framework also allows us to scale from high to low frequency realizations and present out-of-sample risk estimates.

More specifically, this paper addresses examines and compares the tail behaviour of market and limit order returns.[1] In our study, dealers trade on the Dealing 2000-2 electronic FX broking system and have to decide between market and limit orders for their order transmission mechanism. Our analysis is based on actual transaction data rather than the common application of using indicative quotes with no firm commitment to actually transact. This means that we can determine the true realised risk and return characteristics and provide a comparison of return distributions associated with these different forms of order mechanism.

The article is organised as follows. Section 2 briefly describes the theoretical framework. Section 3 then outlines the transaction data analysed. Section 4 provides tail estimates and associated tail risk estimates. Section 5 concludes.

**2. Extreme Value Modelling:**
We use EVT to examine the tail behaviour of the returns associated with limit and market orders.[2] This approach offers a number of advantages for the issue at hand. First, EVT tells us the distribution we should fit, namely, the heavy-tailed Fréchet distribution. And second, it tells us that the tail behaviour of the Fréchet distribution exhibits a self-similarity property that allows for an easy extension for multi-period horizons using a simple scaling rule.

---

[1] In contrast to the approach taken here order choice can be analysed as a function of execution probabilities (see Hasbrouck, 2007; for a comprehensive treatment) and the execution probabilities for our data set are given in Danielsson and Payne (2002b).
[2] Only salient features of relevance are presented (for further details see Embrechts et al, 1997).

We begin by assuming that order type returns are represented by a random variable, $R$, and belong to the true unknown cumulative probability density function $F(r)$. Let $(M_n)$ be the maxima of $n$ random variables such that $M_n = max\ \{R_1, R_2,..., R_n\}$.

Whilst the exact distribution of order returns is unknown the following property of regular variation at infinity gives us a necessary and sufficient condition for convergence to the heavy-tailed extreme value distribution:

$$\lim_{t \to \infty} \frac{1 - F(tr)}{1 - F(t)} = r^{-\alpha} \qquad (1)$$

For a heavy-tailed distribution the probability that the maximum value exceeds a certain price change, $r$, is given by:[3]

$$1 - F^n(r) \approx ar^{-\alpha} \qquad (2)$$

where $a$ represents the scaling constant and $\alpha$ is the tail index, for $\alpha > 0$, and for $r \to \infty$. The tail index, $\alpha$, measures the degree of tail heaviness and also measures the number of bounded moments.

The asymptotic expression is for any given frequency and it is easy to extend the framework to lower frequencies using an $\alpha$-root scaling law (for an application see Cotter, 2007). For instance, taking the single period price changes, $R$, and extending these to a multi-period setting, $kR$, relies on the additivity property of a fat-tailed distribution from Feller's theorem (Feller, 1971, VIII.8):

$$1 - F^n(kr) \approx kar^{-\alpha} \qquad (3)$$

Importantly the tail index, $\alpha$, remains invariant to the aggregation process and does not require estimation of further parameters. Estimation efficiency of $\alpha$ is highest for the highest frequency data thereby giving empirical benefits (Dacorogna et al, 2001a). Moreover, low frequency estimation suffers from a limited sample size. Thus scaling allows us to obtain accurate lower frequency results that otherwise would be difficult to obtain due to sample size considerations.

---

[3] The theoretical framework is presented for upper tail statistics following convention.

The risk levels associated with order type require initial estimation of the tail index and we choose to use the Hill (1975) moment estimator. The Hill estimator represents a maximum likelihood estimator of the inverse of the tail index:

$$\gamma(m) = 1/\alpha = (1/m) \sum [\log r_{(n+1-i)} - \log r_{(n-m)}] \quad \text{for } i = 1....m \quad (4)$$

This tail estimator is known to be asymptotically normal $(0, \gamma^2)$ for $m$ increasing rapidly with $n$. However, there remains the issue of how to determine the optimal tail threshold, $m$ and there has been much debate on this issue (for example see Dacorogna et al, 2001b; and Blum and Dacorogna, 2003). The Hill estimator is asymptotically unbiased but suffers from small sample bias which limits its application to many financial situations. A common applied approach is to use bootstrap methods. For example Danielsson et al (2001) develop a sub-sample bootstrap where the optimal threshold, $m$, is determined from analysis of independently drawn sub-samples. However, Dacorogna et al (2001b) note that this method depends on having a sufficiently large sample size to begin with prior to choosing representative sub-samples. To address the small finite sample problem, we make use of the Huisman et al (2001) weighted least squares regression approach to provide an alternative estimate of the tail index. The approach focuses on obtaining small sample tail-index estimates and are close to being unbiased for simulation of samples as small as 100. The regression exploits the fact that the bias in the tail index estimate can be described as a linear function. Here, a number ($n/2$) of Hill tail estimates are measured and are regressed against the associated thresholds, $m$.

Extreme limit and market returns are compared to determine the quantile or risk levels that would be incurred for various probabilities, $p$:

$$r_p = r_m(m/np)^{1/\alpha} \quad (5)$$

This provides a single period estimate of risk levels for a given horizon at a given probability level, $p$. We can examine both in-sample, $p \geq 1/n$, and out-of-sample, $p < 1/n$, probability levels. The quantile estimate can be scaled using Feller's theorem to provide an alternative way to estimate the tail risks inherent in limit and market orders over lower frequencies. This extreme multi-period risk *(MPR)* measure determines the risk level for a relatively long holding period using the $\alpha$-root scaling law as given below

$$MPR = r_p k^{1/\alpha} \quad (6)$$

One will note that the use of this rule does not require estimation of further parameters.

**3. Data considerations:**

The spot foreign exchange market is the largest spot market in the world and USD/DEM exchange rate was the most actively traded for the period of analysis of this study, namely, the week of the 6th to the 10th of October 1997 (BIS, 1998). The foreign exchange data employed in this study is from the D2000–2 electronic FX broking system run by Reuters. This unique brokerage system is one of the two main electronic brokers, the other operated by the EBS partnership. Unlike other systems dealing only with single dealers, D2000-2 allows for an examination of the activities of multiple dealers. This allows the D2000-2 system to provide a comprehensive description of limit and market order related information. Furthermore, by using actual transaction data, analysis of the D2000-2 data set overcomes problems associated with the use of indicative quotes, which can be misleading because they are not binding trade commitments.[4]

The introduction of electronic broking systems has led to rapid growth in the levels of brokered trade completed electronically. Estimates for the two electronic brokerage systems suggest approximately 40% of all trades were carried by EBS and D2000-2 for the October 1997 week considered in this study. The proportion of trades carried out electronically then grew to 55% in April 2004 (BIS 2004). About the same time, the proportion of inter-dealer business completed by electronic systems had grown to 67% (Bank of England, 2004).

The D2000-2 data set contains comprehensive information on all trading activity in USD/DEM exchange rate for the trading week covering the 6th to the 10th of October 1997.[5] This amounts to 130,535 entries in total. Over 100,000 of these incorporate limit order information and the remainder refer to market orders. The average size of the former order is $2m whereas the average size of the latter is $3m. Thus, limit orders are a

---

[4] Further shortcomings in applying these indicative quotes in microstructure studies include the lack of traded volume information, no details on the timescale of these quotes, and identification of the exit of these quotes not being possible (Danielsson and Payne, 2002a).
[5] For each trading entry, there are ten fields of information detailing the type of event to which it refers, timestamps within one hundredth of a second, price and quantities.

more popular method of trading with higher liquidity, but market orders have larger trades.[6]

The D2000-2 user has access to the best limit buy and sell prices, plus the quantities available at these prices and a record of recent transaction activity. Only filled orders are analysed which occurs in one case in three for limit orders, but always occurs either partially or fully for market orders. This study employs the best bid and offer quotes at the end of each trading interval measured in calendar time to obtain mid-quotes. A sampling frequency of 20 seconds was employed to convert to calendar time. This sampling frequency was chosen to satisfy two requirements, namely to ensure that the information usage inherent in the data set is well utilized, on the one hand, and to try to ensure that intervals did not suffer from a thin trading bias, on the other. Following Andersen et al (2001), all findings are presented for 5-minute intervals to minimize the serial correlation in the bid-ask spread returns induced by non-synchronous trading

To get a preliminary understanding of the return characteristics for limit and market order returns, Table 1 presents some summary statistics. Returns are calculated as the first difference of log prices and volatility is proxied by absolute returns. In general, we obtain the usual stylized facts of excess (positive) skewness and excess kurtosis (and hence non-normality) for both order types. Average returns are negative for market orders but positive for limit orders. However, the standard deviation of returns and the average of absolute realisations clearly indicate that volatility is larger for limit orders. Limit orders also have the propensity for larger extreme values. We find negligible dependence in returns that increase substantially for the volatility proxies in line with financial time series. Here non-synchronous trading induces large negative autocorrelation in the first lag but otherwise results in a lack of significant autocorrelation for the returns series. In contrast, almost half of the lags of market order absolute returns exhibit strong dependency.

---

[6] Furthermore, information on limit orders contain details with timestamps for entry and exit times, a buy/sell indicator, quantity available, quantity traded and price; whereas for market orders, the quantity transacted and price, a timestamp and whether buyer or seller initiated is given (for further details see Danielsson and Payne, 2002a).

INSERT TABLE 1 HERE

**4. Empirical Findings:**

We now turn to the main findings of the study and examine the tail risks of market and limit orders. The tail index estimates and associated standard errors are presented in Table 2. There are similarities in the tail behaviour between extreme limit and market order returns but there are also some notable differences. First, the existence of the heavy-tailed characteristic for limit and market orders is confirmed for all trading periods from the significant t-statistic, $H_o: \gamma = 0$. We also determine the number of defined moments that exist with the hypotheses, $H_o: \gamma = 2$, and $H_o: \gamma = 4$ determining whether there exists $2^{nd}$ and $4^{th}$ moments respectively. Support for a defined variance is never rejected and this finding lends support for the use of the $\alpha$-root scaling law. On the other hand, no evidence is available to support the existence of the $4^{th}$ moment of the underlying distribution, and this is the case for both order types. This finding means that we can reject the hypothesis that either market or limit order returns are normally distributed.

INSERT TABLE 2 HERE

However, the results also show that tail index estimates are consistently lower for limit orders. Since a lower tail index implies a heavier tail density, this indicates that limit orders have heavier tails than market orders. Stability tests to test for significant differences between market and limit order tail values generally support the hypothesis of distinct tail estimates by order type. For example, in both the upper and common tail cases the estimates for the limit order are significantly lower than market orders and these results indicate that the former have significantly heavier tails.

Some extreme risk quantile estimates are presented in Table 3. For example, the first risk level with a probability $p = 2/n$ suggests a 5-minute extreme return that occurs twice in the sample week. The risk levels show that limit orders have larger extreme returns than market orders. For instance, suppose a dealer takes a short trading position, the associated risk that would occur with a probability of $p = 2/n$, and occurring twice in a week's trading is 0.553% for a limit order but only the 0.239% for a market order. To illustrate the size of the risk, if we take an average order ($3m for market and $2m for limit), then the dollars at risk are $11,060 for limit orders and only $7,170 for market orders over 5-minute holding periods.[7] These dollar losses would occur twice over a week's trading.

INSERT TABLE 3 HERE

We also present out-of-sample estimates that would not be available from analysis of the full empirical distribution, *n*. For our dataset for instance $p = 1/2n$ represents a single 5-minute risk estimate that occurs once in every two weeks of trading. Clearly the divergences between order type becomes more pronounced for lower probability levels (eg. $p = 1/2n$), with the risk quantile associated with the upper tail limit order being 1.008% compared to 0.396% for market orders.

Thus far we have assumed that trading for the forex dealers has occurred at the same frequency as our 5-minute sample interval. However, foreign exchange activity may involve trading at lower frequency intervals such as hourly or daily periods. The extreme value estimates we have obtained earlier can be extended to longer intervals using the *α-root* scaling law. Accordingly, Table 3 shows the implied risk levels for a dealer who holds a position for 1 hour, 4 hours and 8 hours respectively. The risk levels are estimated using the 5-minute Hill estimates for the probability, $p = 1/n$, scaled by the extreme value multiplication factor. The results clearly show the potential for limit orders to generate relatively riskier positions over market orders for longer holding periods. For an hour's horizon, the risk level for a dealer's long position of the limit order is 2.067% compared to 0.752% for a market order. These differences also exist for common and upper tail values. In addition the differences increase as the dealer's

---

[7] Obviously divergence of risk levels between order type would be even greater for trades of the same magnitude.

holding period increases: for example, for a market order over 8 hours (approximately a daily interval) on a long position, the limit order has a risk level of 4.848% whereas the market order has a risk level of only 1.588%

**5. Conclusion**

This paper has examined the tail behaviour associated with market and limit order returns. We analyse a unique data set of FX exchange rate transactions to show the true realised risk and return characteristics associated with each type of trade. Using Extreme Value Theory we show the extent to which to which the return distributions associated with each type of trade differ, both for in-sample and out-of-sample contexts and also over different trading frequencies. Overall, the paper finds that whilst limit orders may offer dealers a number of attractions such as increased liquidity, they also exhibit heavier tail behaviour and this results in larger risk levels.

Table 1: Summary statistics for 5 minute limit and market orders

|  | Returns | | Volatility | |
|---|---|---|---|---|
|  | **Limit** | **Market** | **Limit** | **Market** |
| Mean | 1.81E-08 | -1.55E-06 | 4.29E-04 | 2.28E-04 |
| Standard Deviation | 7.86E-04 | 5.00E-04 | 6.59E-04 | 4.45E-04 |
| Skewness | 1.31E-01 | 1.08E+00 | 3.94E+00 | 7.31E+00 |
| Kurtosis | 1.66E+01 | 5.95E+01 | 2.38E+01 | 7.99E+01 |
| No. ACF | 4 | 10 | 10 | 48 |
| Normality | 1.55E-01 | 2.05E-01 | 2.58E-01 | 3.04E-01 |

Notes: With the exception of skewness and kurtosis coefficients, all values are expressed in percentage form. Dependence is outlined with the number of significant ACF coefficients over 100 lags. Normality is formally examined with the Kolmogorov-Smirnov test. The skewness, kurtosis and normality coefficients are all significant at the 5 percent level.

Table 2: Tail index estimates for 5 minute limit and market order returns

| | $\gamma$ | $\gamma=0$ | $\gamma=2$ | $\gamma=4$ | *Stability test* |
|---|---|---|---|---|---|
| **Common Tail** | | | | | |
| Limit | 2.46 | 5.67 | 1.06 | -3.55 | -4.29 |
| | (0.22) | | | | |
| Market | 2.72 | 5.72 | 1.52 | -2.68 | |
| | (0.24) | | | | |
| **Lower Tail** | | | | | |
| Limit | 2.30 | 4.33 | 0.57 | -3.19 | -0.88 |
| | (0.27) | | | | |
| Market | 2.35 | 3.77 | 0.56 | -2.66 | |
| | (0.32) | | | | |
| **Upper Tail** | | | | | |
| Limit | 2.44 | 5.30 | 0.96 | -3.39 | -5.42 |
| | (0.24) | | | | |
| Market | 2.78 | 4.93 | 1.38 | -2.17 | |
| | (0.29) | | | | |

Tail estimates are calculated for common, lower and upper tails. Standard errors are presented in parenthesis for each tail value. Tail estimates are compared to values of 0, 2 and 4 with a critical value of 1.64. The Hill tail estimates are compared across order type (eg. market and limit orders for upper tail) using the stability test outlined by Koedijk and Kool (1992) with a critical value of 1.96.

Table 3: Risk levels for limit and market order returns

|  | Single period | | | Multi-period | | |
|---|---|---|---|---|---|---|
|  | $p = 2/n$ | $p = 1/n$ | $p = 1/2n$ | $k = 12$ | $k = 48$ | $k = 96$ |
| Common Tail | | | | | | |
| Limit | 0.528 | 0.700 | 0.926 | 1.921 | 3.375 | 4.473 |
| Market | 0.253 | 0.329 | 0.427 | 0.820 | 1.365 | 1.761 |
| Lower Tail | | | | | | |
| Limit | 0.554 | 0.749 | 1.012 | 2.207 | 4.032 | 5.45 |
| Market | 0.289 | 0.388 | 0.528 | 1.116 | 2.013 | 2.703 |
| Upper Tail | | | | | | |
| Limit | 0.553 | 0.747 | 1.008 | 2.067 | 3.649 | 4.848 |
| Market | 0.239 | 0.308 | 0.396 | 0.752 | 1.238 | 1.588 |

The estimates are presented in percentage form. The estimates are obtained using the linearly interpolated number of tail values associated with the modified Hill estimator following Huisman et al (2001). The single period estimates are given for various probability levels, from in-sample estimates, for example $p = 1/n$, to out-of-sample estimates, $p = 1/2n$. The multi-period estimates are given for different holding periods, $k$, and for the probability, $p=1/n$.